\documentclass[12pt]{iopart}

\usepackage{graphicx}
\usepackage{iopams}  
\begin{document}

\title[]{Quantum oscillations and subband properties of the two-dimensional electron gas at the LaAlO$_3$/SrTiO$_3$ interface}

\author{A McCollam$^1$, S Wenderich$^2$, M K Kruize$^2$, V K Guduru$^1$, H J A Molegraaf$^2$, M Huijben$^2$, G Koster$^2$, D H A Blank$^2$, G Rijnders$^2$, A Brinkman$^2$, H Hilgenkamp$^2$, U Zeitler$^1$ and J C Maan$^1$}

\address{$^1$High Field Magnet Laboratory, Institute for Molecules and Materials, Radboud University Nijmegen, 6525 ED Nijmegen, The Netherlands.}
\address{$^2$Faculty of Science and Technology and MESA+ Institute for Nanotechnology, University of Twente, 7500 AE Enschede, The Netherlands.}

\ead{A.McCollam@science.ru.nl}

\begin{abstract}
We have performed high field magnetotransport measurements 
to investigate the interface electron gas in 
LaAlO$_3$/SrTiO$_3$ heterostructures.
Shubnikov-de Haas oscillations reveal several 2D conduction subbands 
with carrier effective masses between 1 and 3~$m_e$, quantum 
mobilities of order $3000$~cm$^2$/V~s, and 
band edges only a few millielectronvolts below the Fermi energy.
Measurements in tilted magnetic fields  
confirm the 2D character of the electron gas, and
show evidence of inter-subband scattering.
\end{abstract}

%\pacs{73.20.-r, 71.18.+y, 73.21.-b, 73.40.-c}
\maketitle

\section{Introduction}

It is now well-established that a two-dimensional
electron gas (2DEG) can exist at the interface
between perovskite oxides LaAlO$_3$ (LAO) and SrTiO$_3$ (STO)
\cite{Ohtomo_Hwang, Mannhart_MRS}.
The 2DEG is nominally similar to those 
in semiconductor heterostructures, but supports additional 
phases, such as superconductivity and magnetism,
which are not observed in conventional 2D electron systems, 
and which have great fundamental and technological interest
\cite{Reyren,Brinkman}.
The mechanism of formation of this oxide 2DEG, however,
is not established, and although 
there is a general consensus that
the charge carriers occupy STO conduction bands
modified by the presence of the interface 
\cite{Mannhart_MRS,Chen}, 
the origin and density of the 2D carriers, 
and the details of the electronic bandstructure 
at the interface are not yet fully understood.

Three main scenarios 
have emerged as the possible source of conduction electrons:
\lq electronic reconstruction\rq,~where a build-up of electric 
potential across the polar LAO layers is avoided by 
charge transfer from LAO to STO 
\cite{Okamoto,Mannhart_MRS}; 
donation of electrons by oxygen vacancies in the heterostructure 
\cite{Kalabukhov,Siemons,Herranz}; 
and cationic intermixing and disorder across the interface 
\cite{Willmott,Kalabukhov_PRL}. 
The relative contributions of these
mechanisms seem to be determined by factors 
such as the sample growth conditions, the LAO layer thickness,
and the overall integrity of the crystal structure. 
Electronic reconstruction is widely 
proposed to be the dominant
mechanism in {\em{intrinsic}} samples, where defects, impurities
and disorder are minimised,
but experiments have identified a number of inconsistencies 
in this simple picture 
\cite{Chen, Huijben_AdvMat,Yoshimatsu,Segal,Takizawa,Singh-Bhalla},
and the origin of the conduction electrons and 
the nature of 
their confinement at the interface
are still unresolved questions.

Quantum oscillations in the transport 
(or thermodynamic) properties of metallic and semiconducting materials 
arise directly from the magnetic field dependence of the conduction 
electron energies, and are, therefore, 
a powerful experimental probe of the electronic 
bandstructure close to the Fermi energy  
\cite{Shoenberg}.
They also provide band-specific details of
conduction electron properties such as effective mass and mobility.
Access to this information in LAO/STO 
is highly desirable,
as the bandstructure of the 2DEG can
be expected to reflect the mechanism of its formation, as well
as giving further insight into its behaviour.

Previous quantum oscillation experiments on LAO/STO 
\cite{Herranz,Caviglia,BenShalom_105} 
measured the Shubnikov-de Haas (SdH) effect in the resistivity, 
and could clearly resolve only
a single oscillation frequency corresponding to a single conduction 
band with high mobility charge carriers.
In samples with high carrier density 
(of order $10^{16}$~cm$^{-2}$) \cite{Herranz} 
the oscillations were independent of the magnetic field direction, 
and indicated a three-dimensional Fermi surface containing 
all of the charge carriers. 
These samples were grown under conditions of low oxygen
partial pressure ($< 10^{-5}$ mbar), 
and the high carrier density and mobility 
were believed to arise from uniform doping of the 
STO substrate by oxygen vacancies. 
In references 19 and 20, measurements were carried out on samples
grown or annealed under higher oxygen pressures, with carrier densities
of the order of $10^{13}$~cm$^{-2}$, thought to be 
characteristic of intrinsic samples with few or no oxygen vacancies.
These groups report slightly different values for the
SdH frequency but, in both cases,
the observed conduction band is two-dimensional, with 
very low carrier density 
compared to the
total carrier density extracted from the Hall effect
(of order 20$\%$).
These results raised interesting questions 
about the presence of conduction channels which do not 
contribute to the SdH effect,
or the possibility of multiple valley and spin degeneracies 
\cite{Caviglia,BenShalom_105}, 
and suggested that the LAO/STO conduction-bandstructure is 
considerably more complex than implied by the 
single observed SdH frequency.
A complex bandstructure is also predicted by 
density-functional calculations, which give
a large number of subbands, with quite different carrier 
properties, crossing the Fermi energy
\cite{Popovic,Son,Delugas}.

In this work, we present 
a detailed magnetotransport investigation of 
LAO/STO-based heterostructures,
as a function of temperature and in a range of magnetic 
field orientations, 
from perpendicular to parallel to the oxide layers.
By using very high magnetic fields and high-mobility 2DEGs, 
we have been able to measure Shubnikov-de Haas oscillations
with significantly better resolution than 
previously possible, and have identified and characterised 
several 2D conduction subbands.
The subbands are 
separated by a few millielectronvolts,
and have different 
effective masses and mobilities. 
We find a total carrier density of $\sim 10^{13}$~cm$^{-2}$
contributing to the SdH effect.

\section{Experiments and results}

\begin{figure}
\begin{center}
\includegraphics{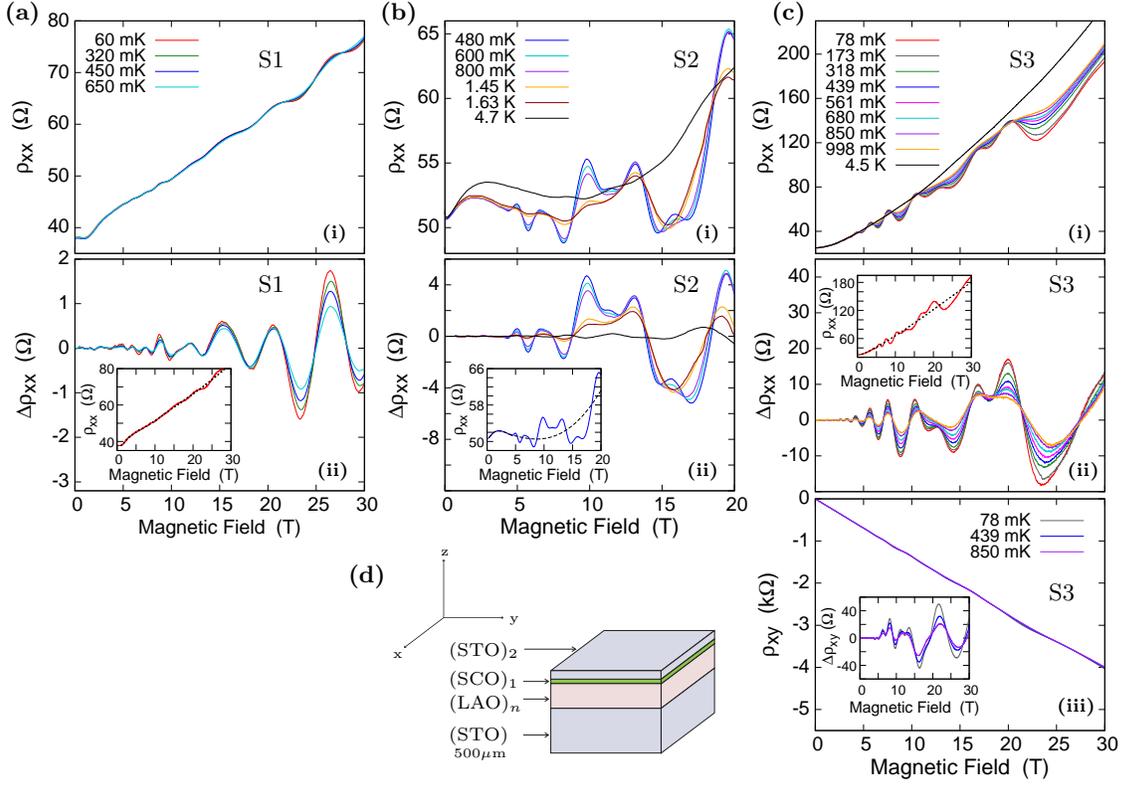}
\caption{(a)--(c)(i): Temperature dependence of 
$\rho_{xx}$ in samples S1, S2 and S3.
(a)--(c)(ii): The oscillations in (i) with background
removed. An example of the background subtracted in each case
is shown as a dashed line in the insets.
(c)(iii): Hall effect of sample S3. The inset shows the
quantum oscillations after subtraction of the linear background.
(d): Sketch of the STO/SCO/LAO/STO structure.}
\label{data}
\end{center}
\end{figure}

We have measured
three heterostructures (labelled S1, S2 and S3),
with the same basic structure, but 
differing slightly in the number of LAO layers  
or in the oxygen partial pressure during growth. 
The samples were grown by pulsed laser deposition,
with ten (S1, S3) or nine (S2) 
monolayers of LAO deposited on a TiO$_2$-terminated STO(001) substrate. 
A single monolayer of SrCuO$_2$ (SCO) and two monolayers 
of STO were grown as capping layers on top of the LAO, 
as illustrated in Fig.~\ref{data}(d).
Oxygen partial pressure of $2 \times 10^{-3}$~mbar was used during
LAO growth for S1 and S2, and this was reduced to  
$1 \times 10^{-5}$~mbar for S3. 
Full details of the growth procedure and parameters
are given as Supplementary Information.

The role of the SCO layer is, at present, not fully understood,
but we find that it considerably increases the mobility of the 
LAO/STO interface (compared to samples prepared in an identical way,
but without SCO) 
\cite{Huijben_condmat}.
We find no evidence of a conducting channel at isolated SCO/LAO or
SCO/STO interfaces, and assume that the single monolayer of SCO 
in our heterostructures does not contribute directly to the conductivity by
supporting an additional 2DEG.
Rather, the SCO layer is thought to indirectly enhance the 
mean free path of mobile electrons at the LAO/STO interface 
by somehow reducing the density of defect donor states
in the structure. 
A detailed investigation of the effect of the SCO layer is presented 
elsewhere 
\cite{Huijben_condmat}.

Our samples had dimensions of 5~mm $\times$ 5~mm in the $xy$ plane,
with electrical contacts made by wire bonding 
through the top surface in a van der Pauw geometry. 
Resistance was measured 
using a standard ac technique with excitation currents
of 1 or 2 $\mu$A.
Experiments were carried out in a dilution refrigerator
and a pumped $^3$He system with a rotatable sample stage.

%Results and Data

Fig.~\ref{data}(a)(i)--(c)(i) shows the longitudinal 
magnetoresistance $\rho_{xx}$ of our samples at various 
temperatures, for magnetic fields aligned perpendicular 
to the $xy$ plane. 
Clear oscillations with strongly temperature-dependent
amplitudes are apparent in each case. 
The complex periodicities indicate that several 
frequencies are superposed.
Fig.~\ref{data}(a)(ii)--(c)(ii) show the same data as in (i),
but with a smooth background removed. An example of the background
subtracted for each sample is shown as a dashed line in the insets of 
these figures.

The Hall effect $\rho_{xy}$ 
of each sample showed quantum oscillations 
similar to those in $\rho_{xx}$, on a large, approximately 
linear background. 
Fig.~\ref{data}(c)(iii) shows data from S3. 
Apart from the oscillation amplitudes, 
$\rho_{xy}$ is
independent of temperature. By performing all measurements 
with both positive and negative magnetic field polarities,
we exclude the possibility 
that the oscillatory Hall effect arises from 
an admixture of $\rho_{xx}$ in $\rho_{xy}$,
or vice versa.
Observable quantum oscillations in 
the Hall effect are relatively unusual, 
but arise from the same Landau level-related phenomena 
as SdH oscillations in
$\rho_{xx}$ \cite{Shoenberg}.
Previous observations of this effect 
have been well documented,
for example, 
in semi-metals
\cite{Reynolds}
and, recently, 
as the first confirmed example
of quantum oscillations in the high temperature 
cuprate superconductors 
\cite{Doiron_Leyraud}.

\begin{figure}
\begin{center}
\includegraphics{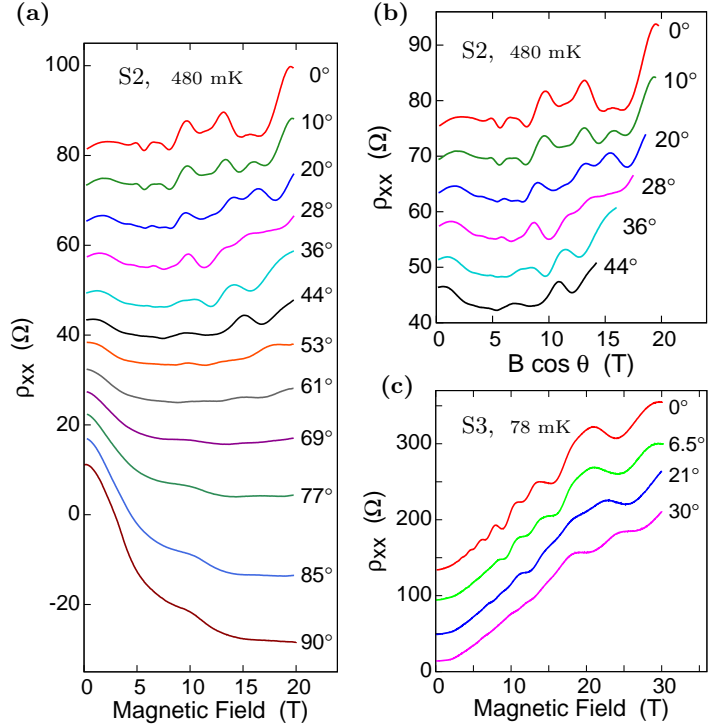}
\caption{Dependence of SdH oscillations on magnetic
field orientation. The angle $\theta$ between the magnetic field 
direction and the normal to the interface is shown beside each of the 
curves. (a) SdH oscillations at different tilt angles for sample S2,  
(b) SdH oscillations for sample S2 
up to $44^{\circ}$ (from (a)), as a function of 
the perpendicular field component $B \cos{\theta}$.  
(c) SdH oscillations at different tilt angles for sample S3.
The curves are offset vertically for clarity. }
\label{angle_data}
\end{center}
\end{figure}

Fig.~\ref{angle_data} shows the
dependence of $\rho_{xx}$ on the angle $\theta$ between the 
magnetic field direction and the $z$-axis of the sample 
for S2 and S3 
(data as a function of magnetic field orientation are 
not available for S1). 
The oscillations are rapidly suppressed, with a change in periodicity, 
as the magnetic field is tilted away from 0$^{\circ}$. 
Over the wider range of angles measured for S2, we see the background 
magnetoresistance change from positive to negative,
and observe the development of a pronounced feature in parallel 
field ($\theta=90^{\circ}$) at $\sim11$~T.

\section{Data analysis and discussion}

We first focus on analysis of the SdH oscillations in $\rho_{xx}$ 
at $\theta = 0^{\circ}$, as shown in Fig.~\ref{data}.
These oscillations have amplitudes
of the order of 10\% 
of the total resistance, which suggests that Landau levels
are not fully resolved in these systems, 
i.e. there is considerable overlap of levels.
In this situation, SdH oscillations are sinusoidal
in inverse magnetic field $1/B$, and are described by 
the general expression 
\cite{LifshitzKosevich,Shoenberg}
\begin{equation}
\Delta \rho_{xx} \propto \sum_i \exp \ (-\alpha_i T_{D_i})\  \frac{\alpha_i T}{\sinh (\alpha_i T)} \ \sin \left( \ \frac{2 \pi p f_i}{B} + \phi_i \right),
\label{LK}
\end{equation}
where $T$ is temperature, and $f$ and  
$\phi$ are the frequency and phase of the oscillations; 
$p$ is the harmonic number.
The amplitude factors contain the term  
$\alpha=2 \pi^2 p k_Bm^*  /\hbar eB$, and  
allow the effective mass $m^*$, and Dingle temperature $T_D$ 
of the charge carriers to be extracted from 
the temperature and magnetic field dependence of the signal
\cite{Shoenberg}.
$T_D$ is a measure of the quantum mobility~$\mu$, 
and can be expressed as $T_D = (e \hbar /2\pi k_B)(1/ m^* \mu)$.

\begin{figure}
\begin{center}
\includegraphics{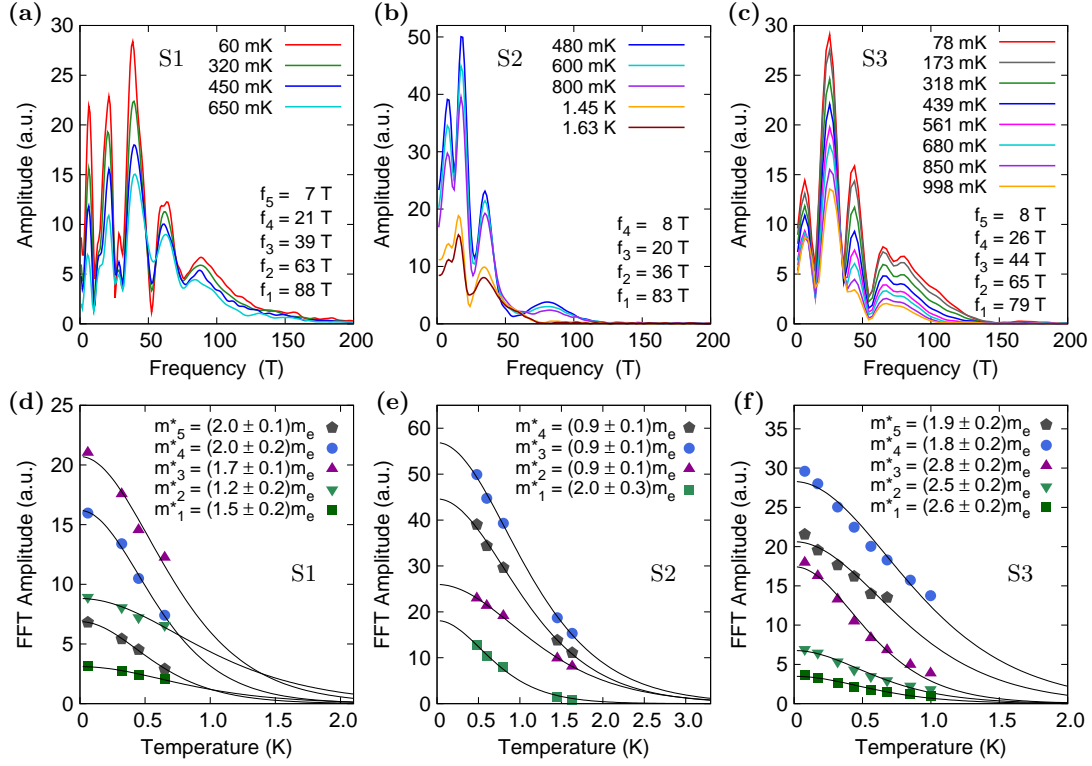}
\caption{(a) Fourier transform between 4 and 30~T for sample S1.
(b) Fourier transform between 4 and 20~T for S2.
(c) Fourier transform between 4 and 30~T for S3.
(d)--(f) Temperature dependence of the oscillation amplitudes
with values of $m^*$ in units of $m_e$.
The solid lines are fits to $\alpha_i T / \sinh(\alpha_i T)$. Some 
of the data are rescaled so that all the data for each sample 
can be shown clearly on a single plot.  
\label{analysis}}
\end{center}
\end{figure}

To analyse the $\rho_{xx}$ data in Fig.~\ref{data}, we subtracted a 
smoothly varying background 
from each of the curves to obtain $\Delta\rho_{xx}$,
as shown, and performed Fourier transforms (FTs) 
in the inverse magnetic field.
Fig.~\ref{analysis}(a)--(c) gives the resulting Fourier spectra for 
the three samples,
and shows that five peaks are visible in the FT for S1 and S3, 
and four for S2, corresponding to five or four distinct
oscillation frequencies in the SdH signals.

The very low frequency oscillations, resolved as FT peaks 
at 7~T or 8~T, may be subject to some 
uncertainty due to the difficulty of achieving perfect background
subtraction. However, in all three samples, 
the background has very weak
or negligible temperature dependence, 
compared to the strong temperature
dependence of the amplitude of the 7 or 8~T peak, 
and none of the background curves we 
subtracted show any periodic component.
We therefore treat the lowest frequency FT peak as a genuine
component of the SdH signal in our samples.
In the absence of harmonic content, which we will discuss below,
the FT results indicate that 
five (S1, S3) or four (S2) high mobility conduction subbands 
contribute to the quantum transport in our samples.
The observation of only four subbands for sample S2 may be 
a result of lower resolution in the FT 
due to the smaller magnetic field range of the measurement
(0 - 20~T for S2, compared to 0 - 30~T for S1 and S3), 
or may be related to the slightly different structure of
S2, which has one fewer LAO layers than S1 and S3.

By fitting the temperature dependence of the FT peak
amplitudes to $\alpha_i T / \sinh(\alpha_i T)$,
we extracted the carrier effective mass in each case
(see Fig.~\ref{analysis}(d)--(f)).
FTs were taken over several magnetic field ranges for each sample, and 
the effective mass extracted for each range.
The  
masses quoted in
Fig.~\ref{analysis}(d)--(f) 
are the mean values of $m^*$ resulting from this procedure.
In all cases, the experimental temperature dependence is 
well-described by the theoretical curves, and 
the errors given reflect the spread in the values of $m^*$
over the full field range, 
rather than the quality of the fits.

If the SdH signal contains harmonics of a given oscillation, both 
the measured frequencies {\em{and}} the temperature dependence 
(we actually extract $pm^*$ from the fits of 
$\alpha_i T / \sinh(\alpha_i T)$)
should scale with the harmonic number $p$. 
This is not the case for any of the 
oscillations we observe in our samples, and we conclude that the
peaks in the FTs correspond to fundamental frequencies, implying 
independent conduction channels.
The frequencies and effective masses obtained from our 
analysis of the SdH oscillations 
are summarised in the first two columns of 
Table \ref{table}. 
We note that analysis of both the conductivity, 
calculated as
$\sigma_{xx} = \rho_{xx}/(\rho_{xx}^2 + \rho_{xy}^2)$, 
and the oscillations in $\rho_{xy}$, gave the 
same results as analysis of $\rho_{xx}$ for a given sample,
within the experimental errors.

\begin{table}
\caption[Table 1]{Subband properties derived from analysis of SdH oscillations in each of the three samples: SdH frequency $f$; effective mass $m^*$; carrier density $n$; subband energy relative to the Fermi energy $E_F-E$; Dingle temperature $T_D$; quantum mobility $\mu$. $^*$~Subband $f_5$ of S1 reaches the quantum limit close to 
10~T, which prevents us from extracting 
a reasonable measure of $T_D$ and hence $\mu$ in this case. 
}
\vspace{0.3cm}
\begin{indented}
\item[]\begin{tabular*}{0.8\textwidth}[b]{@{\extracolsep\fill}lccccc}
\hline
{\rule[0mm]{0mm}{4.5mm}}Sample &$m^*$&$n$ & $E_F-E$ & $T_D$ & $\mu$ \\[0.5ex]
\ \ \ \ \ \ \ \ \ \ \ \ \ \ \ \ \ \ & {\small ($m_e$)} & {\small (10$^{12}$cm$^{-2}$)} & {\small (meV) } & {\small (K)} & {\small (cm$^2$/Vs)}\\[0.5ex]
\hline
{\rule[0mm]{0mm}{4mm}}S1&&&&\\
$f_1$ = 88 T & 1.5(2) &  $4.3$  & 6.9 & 3.55 & 401 \\
$f_2$ = 63 T & 1.2(2) &  $3.0$ & 6.0 & 2.50 & 713 \\
$f_3$ = 39 T & 1.7(1) &  $1.9$ & 2.7 & 0.40 & 3144 \\
$f_4$ = 21 T & 2.0(2) &  $1.0$ & 1.2 & 0.38 & 2813 \\
$f_5$ =\ \ 7 T & 2.0(1) &  $0.3$ & 0.4 & - & -*\\ [0.5ex]
\hline
{\rule[0mm]{0mm}{4mm}}S2&&&&&\\
$f_1$ = 83 T  & 2.0(3)  &   $4.0$  & 4.8 & 1.86 & 558 \\
$f_2$ = 36 T & 0.9(1)  &  $1.7$  & 4.5 & 2.19 & 1085 \\
$f_3$ = 20 T & 0.9(1)  &  $1.0$ & 2.6 & 0.92 & 2609 \\
$f_4$ =\ \ 8 T & 0.9(1)  &  $0.4$ & 1.0 & 2.80 & 821 \\[0.5ex]
\hline
{\rule[0mm]{0mm}{4mm}}S3&&&&&\\
$f_1$ = 79 T  & 2.6(2)  &  $3.8$  & 3.5 & - & - \\
$f_2$ = 65 T & 2.5(2)  &  $3.1$ & 3.0 & - & - \\
$f_3$ = 44 T & 2.8(2)  &  $2.1$ & 2.9 & - & - \\
$f_4$ = 26 T & 1.8(2)  &  $1.3$ & 1.7 & - & - \\
$f_5$ =\ \ 8 T & 1.9(2) &  $0.4$ & 0.5 & - & - \\[0.5ex]
\hline
\label{table}
\end{tabular*}
\end{indented}
\end{table}

\begin{figure}
\begin{center}
\includegraphics{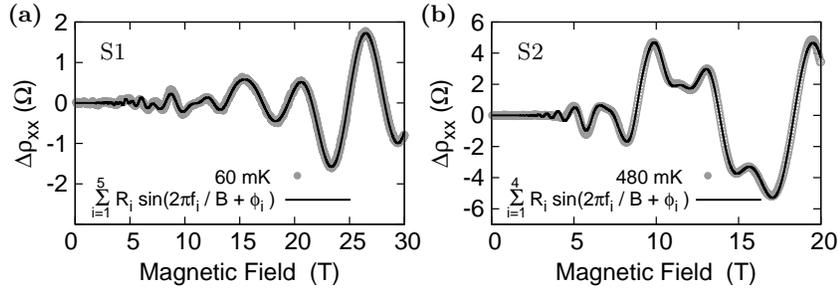}
\caption{Fits of expression (\ref{LK}) to $\Delta\rho_{xx}$ for (a) S1
and (b) S2, using 
$f_i$ and $m^*_i$ from the Fourier transforms.
The grey points show the measured data, and the solid lines represent the 
fits.
\label{LK_fits}}
\end{center}
\end{figure}

As a check of our Fourier analysis, we  
attempted to directly fit the oscillatory magnetoresistance
with expression (\ref{LK}) using the frequencies and values 
of $m^*_i$ extracted from the FTs. 
$T_{D_i}$, $\phi_i$ and an overall 
field-independent amplitude were used as fit parameters.
The fits for S1 and S2 are shown in Fig.~\ref{LK_fits}.
Although the fits are not perfect at very low field, they reproduce 
the data rather well, and suggest that the results
of the FT are reliable. 
We can also use these fits of expression (\ref{LK}) to 
extract $T_D{_i}$, and hence calculate 
approximate values of $\mu$
for each of the subbands in S1 and S2.
These results are given in the final two columns of 
Table \ref{table}.
We emphasise that the values of $T_D{_i}$ and $\mu_i$ are
approximate, but they clearly show that the 
mobilities are different for different subbands. 
Moreover, the highest values of $\mu_i$ in each sample
are in keeping with the magnetic field of $\sim$~5~T at 
which the quantum oscillations begin to appear. 
(This onset field depends broadly on the ratio
of Landau level width to cyclotron energy, and is inversely proportional
to the mobility of the carriers in the relevant subband.)

Although reasonable fits of expression (\ref{LK}) could
be obtained for S1 and S2, it was not possible to achieve even a 
moderately good fit to $\Delta\rho_{xx}$ for S3,
and we were therefore unable to extract values for
$T_D{_i}$ and $\mu_i$.
SdH oscillations in S3 start to become resolvable 
just below 2~T, which indicates that the mobility of at least
one subband in this sample must be considerably higher 
than in S1 or S2.
In the limit of well-resolved Landau levels, with little or no 
overlap between neighbouring levels, SdH oscillations
are still periodic in $1/B$ but are no longer sinusoidal, and the 
LK expression (\ref{LK}) ceases to be valid. 
It is therefore possible 
that the failure of (\ref{LK}) for S3 is related to the higher
mobility of carriers in this sample.

As a brief comment on the possible contribution of spin-splitting
to the SdH signals in our LAO/STO samples, we note that for the usual, 
field-linear Zeeman splitting of the Fermi surface, 
quantum oscillations can only 
provide information about spin-splitting and 
the $g$-factor of a system in certain
specific circumstances: when so-called \lq spin-zeroes\rq~are
observed; when the {\em{absolute}} oscillation amplitudes are
accurately known; or when the signals contain significant harmonic 
content \cite{Shoenberg}.
None of these circumstances apply in our
measurements, and we are not aware of any evidence of non-linear
splitting of the Fermi surface in LAO/STO.
We therefore conclude that spin-splitting 
does not contribute in a resolvable way to the SdH signals 
we observe, so that, even though the Zeeman effect may be relatively 
large, we cannot extract information about
the $g$-factor in our samples.

%%% Now discuss tilted field data %%%

We now consider the magnetotransport in tilted magnetic
fields. Fig.~\ref{angle_data}(a) shows $\rho_{xx}$
for sample S2 as the magnetic field orientation is 
rotated from perpendicular
to parallel to the plane of the LAO/STO interface.
The suppression of the SdH oscillations 
as $\theta$ is increased shows that 
the conduction electrons are confined to the interface plane,
such that quantum oscillations are 
generated only by the perpendicular component of 
the applied field.
In the case of a single occupied subband, 
Landau level separation, 
and hence the period of the SdH oscillations,
would be expected to scale with $1/B \cos \theta$. 
When multiple 2D subbands are
occupied, however, mixing 
of Landau quantisation with the confinement potential
leads to anti-crossings of 
discrete levels from different subbands 
as the field is tilted \cite{Maan}.
Landau level energies therefore show a complicated
magnetic field dependence,
which progressively modifies
the periodicity of SdH oscillations
as a function of $\theta$.
The data in Fig.~\ref{angle_data}(a) show just this behaviour,  
which can be clearly seen as a function of the perpendicular field
component in Fig.~\ref{angle_data}(b),
and confirm both the two-dimensionality of the electron system 
and the contribution to transport of multiple occupied subbands.
We do not have data in tilted magnetic
fields for sample S1, but
data for S3 (Fig.~\ref{angle_data}(c)) show the same 2D, 
multi-subband behaviour.

It can be further seen from
Fig.~\ref{angle_data}(a)  
that, as $\theta$ is increased towards  $90^{\circ}$,  
the magnetoresistance of sample S2 becomes strongly negative 
and develops a significant drop at $\sim11$~T.
Similar behaviour in multi-subband semiconductor 2DEGs 
is attributed to a  decreasing contribution 
of inter-subband scattering to the total resistance,
as the parallel field component
$B_{\parallel}$ shifts the size confinement energies
and depopulates the higher subbands.
The so-called {\em{diamagnetic shift}} is given by
\begin{equation}
\Delta E_n = \frac{e^2B^2_{\parallel}}{2m_n^*}\ \vert <z_n^2> - <z_n>^2\vert, 
\label{diamagnetic_shift}
\end{equation}
where the final term 
is the square of the spread of the
electron wavefunction in the confinement direction
for the $n^{th}$ subband
\cite{Beinvogl}.
When the highest occupied subband is shifted across the 
Fermi energy, it depopulates completely and becomes 
unavailable to scattering processes, leading to a plateau
and subsequent drop in $\rho_{xx}$
\cite{Englert,Portal}.

In order to estimate the diamagnetic shifts expected
in sample S2, we have used a very simple model
of a uniform triangular confining potential at the interface.
The similar masses measured for the three highest subbands 
in S2 suggest that they are split from the same conduction band,
and, based on their energy spacing,
we calculate an approximate value of 
$2.6 \times 10^{-5}$~V/{\AA}
for the electric field confining the charge carriers
in these three highest subbands.
(The details of this calculation are given in 
the Supplementary Information.)
The consequent energy shift would cause the highest subband to 
empty completely at 
$B_{\parallel} \sim 17$~T, 
which is quite close to the drop in resistance we observe at 
$B_{\parallel} \sim 11$~T, 
and supports a description of the magnetoresistance 
in terms of reduction of inter-subband scattering and 
depopulation of the highest subband as $B_{\parallel}$ is increased.
This suggests a very close analogy between the behaviour
of the LAO/STO 2DEG and that of multi-subband semiconductor
2DEGs. 
A significant implication of this observation
is that reducing the number of occupied subbands by
other techniques, for example, via electric field control 
of the carrier density, should also lead to reduced inter-subband 
scattering and thus to higher mobility samples.

For a 2-dimensional electron system with parabolic conduction
bands and circular Fermi surface sections, the  
frequency $f_i$ of quantum oscillations is proportional to 
the carrier density in a given subband $n_i=N_vN_sef_i/h$, where
$N_v$ and $N_s$ are valley and spin degeneracy, respectively.
We use this relation, assuming a single valley and twofold
spin degeneracy, to calculate the carrier density for each of the
subbands we have measured. We also use the corresponding 2D 
density of states to calculate the energies of the subband edges
relative to the Fermi energy, $E_F-E_i =n_i\pi\hbar^2/m_i^*$. 
These values are given in columns 3 and 4 of Table \ref{table}.

The total carrier density contributing to the SdH effect is 
similar for all three samples: 
$n_{SdH} = \sum_i n_i \sim1\times10^{13}$cm$^{-2}$.
This is considerably higher than $n_{SdH}$
measured previously in LAO/STO 
\cite{Caviglia,BenShalom_105}, and is close to the
saturation density generally observed in conducting samples
believed to be free from extrinsic carriers \cite{Thiel}.
We cannot, however, straightforwardly extract values 
for the total carrier densities from our Hall effect data
for comparison, as the Hall coefficient 
is necessarily a combination of contributions from multiple subbands
with different mobilities, and must also comprise the 
effects of inter-subband scattering \cite{Zaremba}.
Naive calculations of $\rho_{xy}$ based only on $n_i$ and $\mu_i$
extracted from our SdH results
show poor agreement with the experimental Hall data 
(see Supplementary Information).

\begin{figure}
\begin{center}
\includegraphics{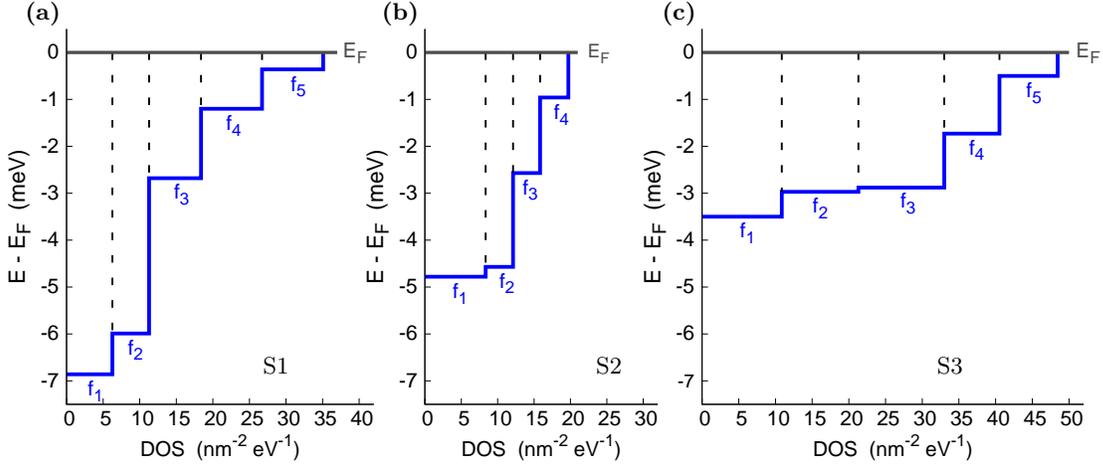}
\caption{Density of states (DOS) for the multi-subband 2DEG in our 
three samples, extracted from
analysis of the SdH oscillations. (a) Sample S1. (b) Sample S2. (c) Sample S3.
\label{subbands}}
\end{center}
\end{figure}

Fig.~\ref{subbands} shows the subband energies $(E_F-E)$
versus density of states for each of our samples. 
The most striking aspect of these subband structures is the 
small energy scale involved. 
The lowest subband edge is only a few millielectronvolts
below the Fermi energy in each sample, and the separation of the 
subbands is mostly $\sim 1$~meV or less. 
We emphasise the observation of this small energy scale,
as bandstructure calculations predict energies of at least an order
of magnitude larger, in the range of tens to hundreds of meV, 
for the occupied conduction subbands in LAO/STO
\cite{Popovic,Son,Delugas}.
The calculated bandstructures assume that a full 
electronic reconstruction takes place at the interface,
with the transfer from LAO to STO of 
0.5 electrons per 2-dimensional unit cell
\cite{Okamoto, Mannhart_MRS}.  
The large discrepancy between our measurement and calculated
subband energies raises questions about both the predicted 
evolution of the conduction bandstructure, and the total 
carrier density possible in the 2DEG.

Although previous measurements on LAO/STO 
could clearly resolve only a single SdH 
frequency \cite{Caviglia,BenShalom_105}, 
and so were unable to provide any information about
the overall subband structure, SdH experiments on 
$\delta$-doped STO \cite{Kozuka, Kim} reveal
frequencies and effective masses in a similar range 
to those we observe in LAO/STO, and 
imply a similar energy scale for the subband structure.
The subbands in electric-field-induced STO 2DEGs \cite{Ueno} 
were also found to be only a few meV below the Fermi energy
for carrier densities close to $1 \times 10^{13}$~cm$^{-2}$.
2DEGs on the cleaved surface of STO, however, 
studied by angle-resolved photoemission (ARPES) 
experiments, have subband structures with band edges
100--200 meV  below the Fermi energy and energy splitting of order
tens to 100 meV
\cite{Santander-Syro,Meevasana}, perhaps suggesting a different
character of the 2DEG in these surface systems.
We note that the subband energies we observe in LAO/STO are at 
the limits of resolution for ARPES and related experiments, 
and may shed light on the failure, so far, 
to resolve the bandstructure at the
LAO/STO interface using these techniques.
The relatively weak confining potential implied by the 
small subband energies may also be reflected in the 
lower than expected core-level shifts observed in photoemission
studies
\cite{Yoshimatsu, Segal, Takizawa}.

Fig.~\ref{subbands} and the results summarised
in Table \ref{table} show that there are small
differences in both the bandstructure and the 
properties of the charge carriers between our three samples.
We attribute this primarily to the differences in 
structure and growth conditions for each of the samples
(see Supplementary Information),
as varying the amount of LAO 
in the heterostructure, or the oxygen pressure during growth,
has been shown to modify the properties of the 
2DEG in several ways  
\cite{Bell,Takizawa, Huijben_AdvMat, Siemons, Herranz}.
Of particular relevance to this work is that the 
number of LAO layers deposited and the 
density of defects such as oxygen vacancies affect the 
electric field which builds up 
across the polar LAO, and can thus be expected to 
directly affect the conduction band energies at the 
interface \cite{Son,Popovic,Singh-Bhalla}.

The different effective masses we observe 
in a given sample are suggestive of subbands with 
different orbital character.
The conduction bandstructure of bulk STO consists
of three degenerate bands with minima at the gamma-point of the 
Brillouin zone, arising from the Ti 3$d\ xy$, $xz$ and $yz$ orbitals.
The degeneracy of these bulk bands is lifted by a low temperature
structural transition and by spin-orbit coupling
\cite{Mattheiss_lowT}, and further splitting  
into 2D subbands is
expected with the introduction of the LAO interface and 
associated confinement potential.
The $d_{xy}$-derived subbands are symmetric in the interface
plane and give circular Fermi surfaces with light masses, whereas
the $d_{xz/yz}$ subbands are expected to be weakly dispersing
along one of the in-plane directions, and should lead to 
elliptical Fermi surfaces and heavy effective masses
\cite{Popovic,Son,Delugas}.
ARPES results on STO surface 2DEGs have yielded values of 
$\sim 0.7~m_e$ and 10--20~$m_e$ for the light and heavy masses,
respectively \cite{Santander-Syro}.
The masses we observe, however, are in a small range 
from $\sim$~1--3~$m_e$, and suggest that in LAO/STO
we measure carriers in hybridised subbands,
with mixed orbital character 
and masses intermediate between light and heavy.
The differences between masses in different subbands should
thus reflect the degree of hybridisation, and can be expected
to depend sensitively on factors such as the detailed bandstructure 
and the spin-orbit coupling in a particular sample.

%%%%%

Even though we have measured three slightly different samples,
the overall consistency of the 
picture we extract from our SdH results 
is notable.
The number of subbands and the values of the 
SdH frequencies are very similar for the three samples, indicating
similar Fermi surfaces
as well as a comparable 
total carrier density in each case. 
The effective masses fall within a small range between 
$\sim$~1 and 3~$m_e$, 
and,
in all samples, the energy gaps between subbands are a few meV or less,
with the lowest occupied subband 
less than $\sim$ 7 meV below the Fermi energy.
The general continuity of our results, despite the 
differences in sample structure and growth conditions,
suggests that we are probing the instrinsic electronic 
behaviour of these LAO/STO 2DEGs. By revealing the
detailed conduction bandstructure and the properties of the 
high mobility charge carriers, the results we have presented provide 
an opportunity to considerably advance our understanding of the 
formation and behaviour of this 2-dimensional 
electron system, and strongly motivate further 
related experimental and theoretical studies.

\section{Summary}

We have shown that multiple 2D subbands, with different effective masses 
and mobilities, are responsible for conduction 
in the LAO/STO 2DEG. 
We observed SdH oscillations 
consistent with those in previous work
\cite{Caviglia, BenShalom_105}, 
but have resolved and characterised
several additional subbands
and a correspondingly higher
density of carriers contributing to quantum transport.
This has allowed us to calculate the 
energies of successive subband edges, and reveals an
energy scale more than an order of magnitude smaller than
predicted by bandstructure calculations.

In tilted magnetic fields we find
evidence of significant inter-subband scattering  
modified by a diamagnetic shift of the subband energies. 
This observation strengthens the analogy between
LAO/STO and the behaviour of semiconductor 2DEGs,
and suggests that reducing the number of
occupied subbands may be a route to  
even higher mobility 2DEGs in this oxide system.

\section*{Acknowledgements}

This work has been supported by the Stichting Fundamenteel
Onderzoek der Materie (FOM) with financial support
from the Nederlandse Organisatie voor Wetenschappelijk
Onderzoek (NWO).
AMC acknowledges support from a 
Marie Curie IIF grant.

\section*{References}

\newpage

\newcommand{\0}{\rm{o}}

\title[]{\Large \bf Supplementary Information}

\section*{Sample growth}

The SrTiO$_3$/SrCuO$_2$/LaAlO$_3$/SrTiO$_3$ heterostructures
used in our study were grown by pulsed laser 
deposition,
with the growth process monitored {\em{in situ}} by reflection high 
energy electron diffraction (RHEED). 
LaAlO$_3$ was deposited on a TiO$_2$-terminated single crystal
SrTiO$_3$(001) substrate at 850~$^{\0}$C in an oxygen
environment. 
Either the oxygen pressure during growth, or the number of
LaAlO$_3$ layers was varied slightly between samples, as follows:\\\\
\begin{tabular}{|c|c|c|}
\hline
{\rule[0mm]{0mm}{4.5mm}}Sample & No. of LaAlO$_3$ layers & O$_2$ pressure during LaAlO$_3$ growth\\[0.5ex]\hline
{\rule[0mm]{0mm}{5mm}}
S1 & 10 & $2 \times 10^{-3}$~mbar \\[0.5ex]\hline
{\rule[0mm]{0mm}{5mm}}
S2 & 9 & $2 \times 10^{-3}$~mbar \\[0.5ex]\hline
{\rule[0mm]{0mm}{5mm}}
S3 & 10 & $1 \times 10^{-5}$~mbar\\[0.5ex]
\hline
\end{tabular}
\\\\\\
The sample was cooled from 850~$^{\0}$C to 600~$^{\0}$C in the LaAlO$_3$
growth pressure at a rate
of 50~$^{\0}$C/min, and 
a single monolayer of SrCuO$_2$ and two monolayers of 
SrTiO$_3$ were then deposited at 600~$^{\0}$C 
in $6 \times 10^{-2}$~mbar of oxygen for all samples.
This part of the growth procedure also acts as an annealing
step for the LaAlO$_3$.
The fluence of the laser pulses was 1.3~J/cm$^2$, the spot size was 
1.76~mm$^2$ and the repetition rate
was 1~Hz. The target-substrate distance was 50~mm.
After growth, the samples were slowly cooled to room temperature
in $6 \times 10^{-2}$~mbar oxygen at a rate of 10~$^{\0}$C/min.
\section*{Hall effect}

In figure 1 we show Hall effect data (solid lines) 
from all three samples.
As described in the main paper, the Hall effect shows quantum oscillations,
but is otherwise temperature independent
and approximately
linear in magnetic field - there is a slight change of slope between 3~T 
and 5~T in samples S1 and S2 .

\begin{figure}[h]
\centering
\includegraphics{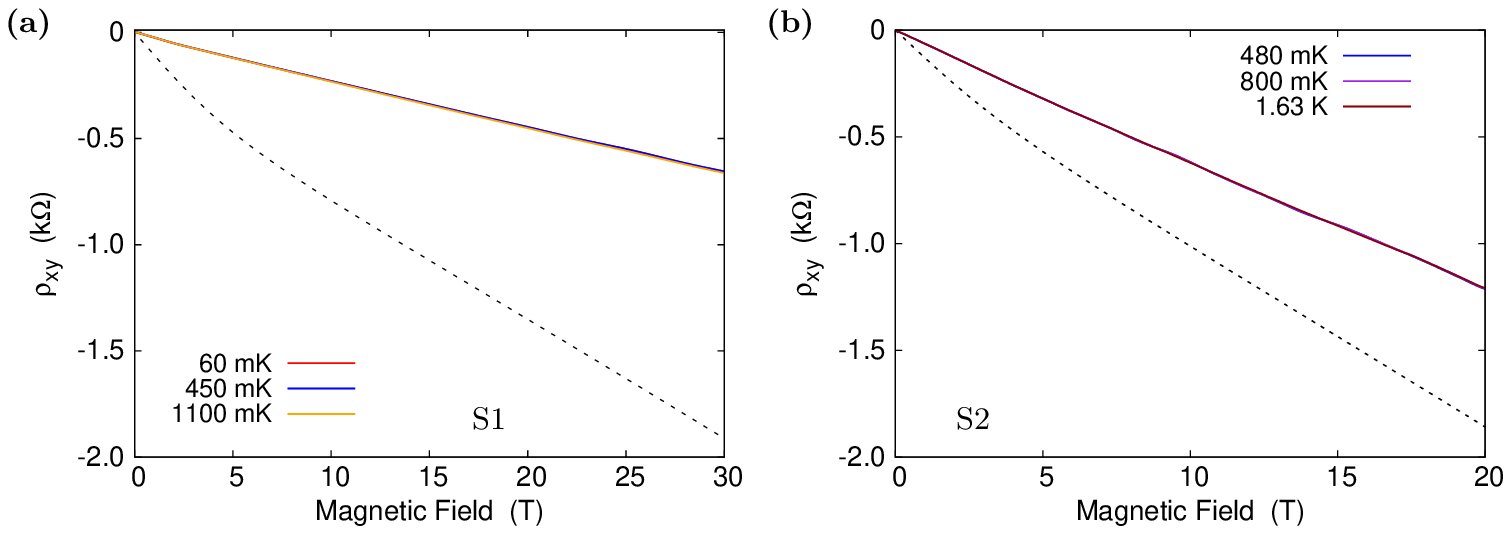}
\begin{minipage}[c]{.49\textwidth}
      \centering
\vspace{0.3cm}
      \includegraphics{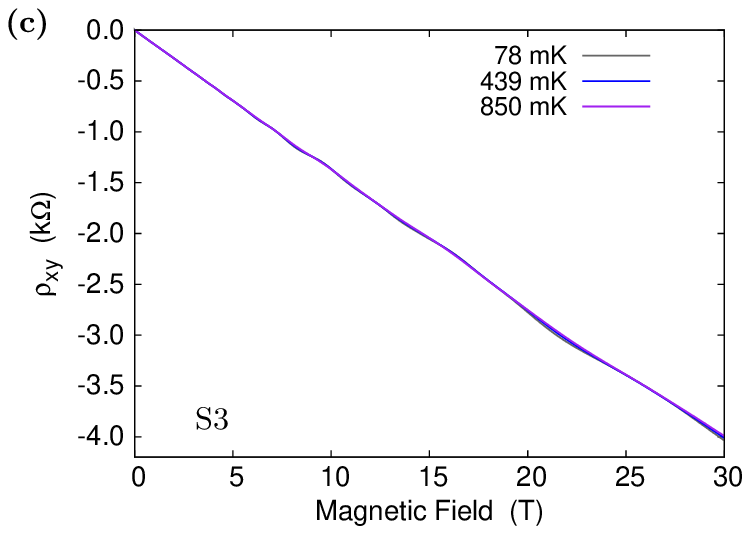}
  \end{minipage}
  \centering
  \begin{minipage}[c]{.49\textwidth}
    {\small{Figure 1: Hall effect at different temperatures for (a) S1,
(b) S2, and (c) S3.} The dotted lines in (a) and (b) show the expected 
Hall effect
for S1 and S2 calculated using the multiband model (1) 
and the mobilities and carrier densities 
for these samples extracted from the 
quantum oscillations.~($\mu_5$ for S1 is estimated, see text.)}
 \end{minipage}
\end{figure}

Using the relaxation-time approximation to the Boltzmann transport equation,
it is straightforward to derive the Hall effect expected in a 
multi-subband system [s1], which can be expressed in terms of the 
carrier densities and 
mobilities of each subband as follows:
\begin{equation*}
\fl
\ \ \ \ \rho_{xy} = \left(\frac{1}{e}\right)
\frac{\left(\frac{n_1 \mu_1^2 }{1+(\mu_1B)^2} + \frac{n_2 \mu_2^2}{1+(\mu_2B)^2} + ...\ \right)}{\left(\frac{n_1 \mu_1}{1+(\mu_1B)^2} + \frac{n_2\mu_2}{1+(\mu_2B)^2} + ...\ \right)^2+ B^2\left(\frac{n_1\mu_1^2}{1+(\mu_1B)^2} + \frac{n_2\mu_2^2}{1+(\mu_2B)^2} + ...\ \right)^2} \ B\ \ \ \ \ (1)
\end{equation*}
\vspace{0.3cm}
The multiband Hall effect has, in general, a complicated dependence on the 
applied magnetic field, but reduces to a more familiar linear field dependence
in the extreme low field/low mobility or high field/high mobility limits:\\
\begin{equation*}
\rho_{xy} = \frac{n_1 \mu_1^2 + n_2 \mu_2^2 + ...}{e(n_1 \mu_1 + n_2\mu_2 + ...)^2}\ B\ \ \ \hspace{3cm} (\mu_iB) << 1\ \ \ \ \ \ \ (2)
\end{equation*}
\begin{equation*}
\rho_{xy} = \frac{1}{e(n_1 + n_2 + ...)}\ B\hspace{4.4cm} (\mu_iB) >> 1\ \ \ \ \ \ \ (3)
\end{equation*}
\vspace{0.3cm}
We note that $\mu_iB \equiv (\omega_c\tau)_i$, where $\omega_{c_i} = eB/m_i^*$ 
is the cyclotron frequency, and $\tau_i$ is the
time between scattering events, or relaxation time, in a given subband.

The wide magnetic field range in our measurements means that the 
full expression (1) for $\rho_{xy}$ should be the relevant one. 
In figures 1(a) and (b) we show, as dotted lines, 
$\rho_{xy}$ calculated using this 
expression with $\mu_i$ and $n_i$ extracted from the SdH data for S1 and S2
(main paper, Table 1). $\mu_5$ for S1 is not available from our measurements,
but is estimated as 1000 cm$^2$/V s.
Due to the low carrier density in this subband, the value of $\mu_5$
has a very weak influence on the shape of the curve.

For sample S3, we do not have values for $\mu_i$ but we can estimate
the behaviour in the high field limit based on expression (3). This gives 
a high field Hall coefficient $R_H = 1/(e n_{SdH}) = -58.3$~m$^3$/C, 
where $n_{SdH} = \sum_i n_i$, and the $n_i$ for S3 are given in 
Table 1 of the main paper.   The slope of the experimental
Hall signal for S3, shown in figure~1(c), is 
considerably {\em{steeper}}, at $R_H = -136.4$~m$^3$/C.

The discrepancies between the
calculated and experimental curves suggest that effects such
as inter-subband scattering [s2],
or the presence of additional occupied subbands 
which do not contribute to the SdH effect, may need to be taken 
into account in any description of the Hall effect in this system.

It is clear from expression (1) that an oscillatory
Hall coefficient can be
expected in a system with multi-band transport, where 
one or more of the bands have an oscillatory conductivity (mobility).
The oscillatory Hall effect therefore has the same origin as the SdH
effect in $\rho_{xx}$.
As can be seen from figure 1, the oscillations we observe in 
our samples are very small 
compared to the large non-oscillatory Hall signal (of order 1\% or less),
but their amplitudes are comparable to the amplitudes
of the oscillations in $\rho_{xx}$ for each sample.

\section*{Modeling the diamagnetic shift in $B_{\parallel}$}

For sample S2 in tilted magnetic fields, we observe a strong negative
magnetoresistance which develops a plateau and downward step
for $B_{\parallel} \sim 11$~T 
(see Fig.~2 of the main paper).
This behaviour typically indicates the reduction of inter-subband
scattering and depopulation of the highest subband due to diamagnetic
shift of the subband energies in a magnetic field parallel
to the 2DEG 
[s3, s4, s5].
The diamagnetic shift expected for electrons in the $n^{th}$
subband is given by
\begin{equation*}
\Delta E_n = \frac{e^2B^2_{\parallel}}{2m_n^*}\ \vert <z_n^2> - <z_n>^2\vert, 
\end{equation*}
which is expression (2) in the main paper.
$\vert<z_n^2> - <z_n>^2\vert$ is the square of the spread of the 
electron wavefunction in the $z$-direction.

In order to try and quantify this effect in our sample, we 
have constructed a very simple model of the 
confinement at the interface based on a band model of 2D
electrons which are free in the $xy$-plane and confined
in the $z$-direction, in analogy to the well-known situation in
semiconductor heterostructures.
The three highest subbands in S2 have similar effective masses of
$m^* \sim 0.9~m_e$ (Table 1, main paper), which suggests that they 
are split from the same bulk conduction band due to confinement
at the interface. 
The energy spacing we have measured for these subbands
implies a triangular confining potential $V(z) = eFz$, where
$F$ is the electric field at the interface.

The wavefunctions in a triangular potential are Airy functions
and their solutions are well known to be 
\begin{equation*}
\nonumber
\zeta_n(z) = \mathrm{Ai} \left(\frac{2 m^*_z eF}{\hbar} \left(z-\frac{E_n}{eF}\right)\right) 
\end{equation*}
[s6, s7], with eigenvalues
\begin{equation*}
\ \ \ \ \ E_n \sim \left(\frac{\hbar^2}{2m^*_z}\right)^{1/3}\left(\frac{3 \pi e F}{2} \left(n + \frac{3}{4}\right)\right)^{2/3}\ \ \ \ \ \ \ \ \ n = 0,1,2, ... .
\end{equation*}
These wavefunctions and eigenvalues  
depend on $m^*_z$, rather than
the in-plane effective mass
$m^*_{xy}$ that we extract from our SdH data.
The lightest charge carriers in LAO/STO
(and in bulk STO) 
are expected to occupy conduction bands with $d_{xy}$ symmetry 
and $m^*_z > m^*_{xy}$. We do not have values of $m^*_z$ for our samples, 
and therefore use a ratio of $m^*_z/m^*_{xy} \sim 2$, based
on SdH results in bulk (Nb doped) STO [s8].
Although the lowest subband in S2
has a heavier effective mass of $2~m_e$,
this is still relatively light, and we assume that this subband also
has predominantly $xy$ character.
For convenience, we label the subbands
($L$) (for $m^* = 0.9 m_e$) and ($H$) (for $m^* = 2 m_e$).  
Using the energy spacing of the three ($L$) subbands 
(column 4 of Table 1 in the main paper), and assuming
$n = 0, 1$ and $2$, we calculate a value of 
$F = 2.6 \times 10^{-5}$~V/{\AA} for S2.

\begin{figure}[t]
\includegraphics{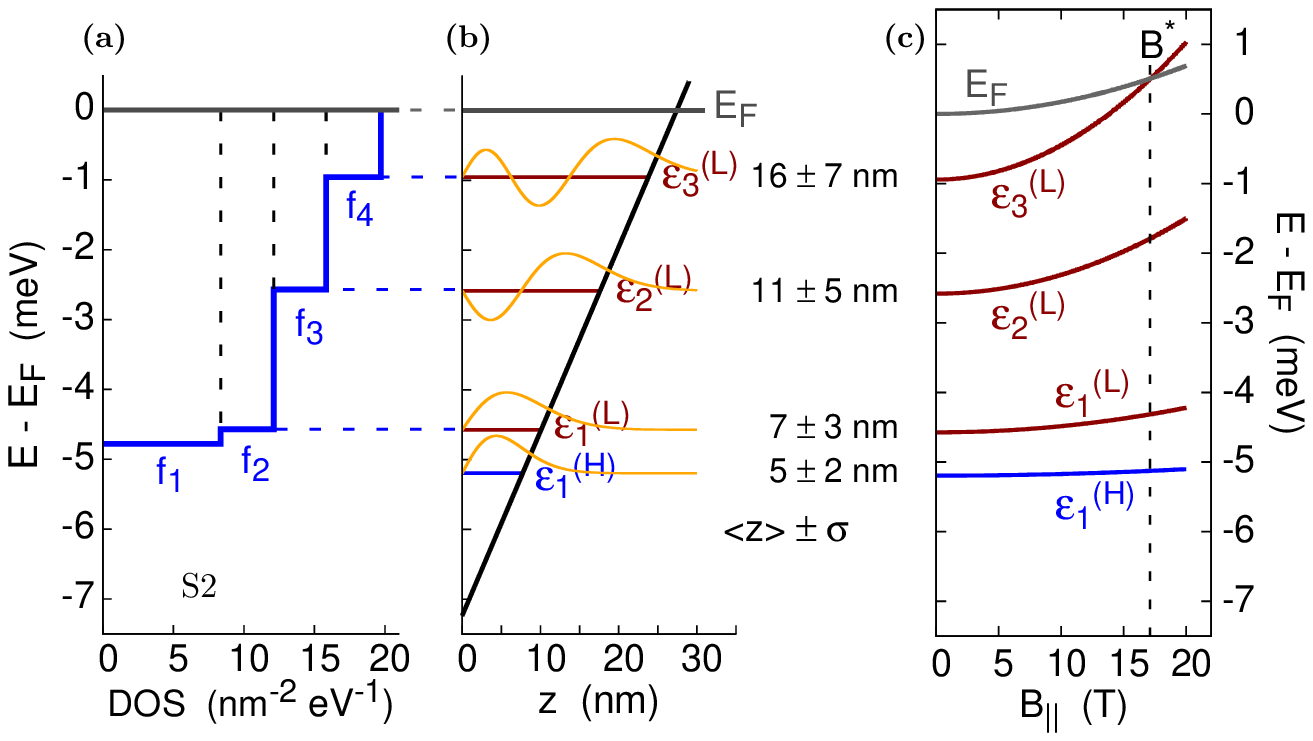}\\
{\small{Figure 2: (a) Subband structure for the 2DEG in sample S2,
derived from analysis of the SdH oscillations.
(b) Potential well structure estimated from the subband
spacing in S2. The wavefunction (in arbitrary units) 
and average $z$-position corresponding to each subband
in this model are also shown;
$\sigma = \sqrt{\vert<z_n^2> - <z_n>^2\vert}$ 
is the 
spatial extent of the wavefunction. 
(c) Evolution of subband energies in the potential well of (b)
due to the expected diamagnetic shift.}}
\end{figure}

The potential well structure described by this model  
is shown in figure~2(b), including sketches
of the wavefunction for each subband. 
We emphasise that this model is approximate, but 
comparison with figure~2(a), which shows the 
measured subband structure for S2 (taken from Fig.~5 in 
the main paper), shows that  
it reproduces the overall spacing of the subbands rather well.
Taking the average values  $<z_n> = 2E_n/3eF$ and 
$<z_n^2> = 6/5 <z_n>^2$ [s6, s7], we then calculate the
diamagnetic shifts expected for each of the subbands, based on the 
potential well in figure~2(b).
Figure~2(c) gives the resulting subband energies as
a function of magnetic field parallel to the
interface, and shows that the highest subband is expected to 
empty completely at $B^* \sim 17$~T.

The value of 
 $B^*\sim17$~T is in quite good agreement with 
the drop in magnetoresistance we observe in sample S2 at 
$B_{\parallel} \sim 11$~T (Fig.~2(a), main paper), 
and supports an explanation of this feature as arising from
depopulation of the highest subband.

In the triangular potential shown in figure~2(b),
the $n =$~1, 2 subbands with $m^*=2~m_e$ 
would also be expected.  
These are not observed, but lower mobilities
for the slightly heavier carriers may explain their absence, as these carriers 
would then contribute only weakly, or not at all, to the 
SdH oscillations. The presence of these subbands would not
significantly affect our estimate of $B^*$, only shifting it
to slightly lower field.

The model potential well in figure~2(b)
also provides an estimate of the 
thickness of the electron gas, and we show the 
average distance of carriers from the interface 
$<z>$ with  the spatial extent of each wavefunction 
at the right hand side of the figure.
These values are consistent with the 
thicknesses of LAO/STO 2DEGs measured by other techniques
[s9, s10], 
particularly when we consider that
80\% of the carriers we observe are accommodated 
in the lowest two subbands.

Finally, we note that there is growing evidence of a large number
of localised charge carriers in the LAO/STO 2DEG
[s11, s12, s13], and that 
these carriers are expected to be more closely confined to 
the interface than the mobile carriers contributing
to transport [s14, s15].
This would suggest a real confining potential which is non-uniform,
and considerably steeper at the bottom of the well,
with the localised carriers occupying lower-lying
subbands than those we observe in the SdH effect.
We do not consider this picture in detail, 
but such a scenario is not expected to 
significantly affect the conclusions we have drawn here about the 
high-mobility carriers we have measured in this LAO/STO system.

\section*{References}

\noindent
$[\mathrm{s}1]$ N W~Ashcroft and N D~Mermin, {\em{Solid State Physics}} (Harcourt Brace College

\hspace{0.15cm}Publishers, 1976).\\
$[\mathrm{s}2]$ E~Zaremba, Phys.\ Rev.\ B {\bf 45}, 14143 (1992).\\
$[\mathrm{s}3]$ W~Beinvogl, A~Kamgar and J F~Koch, Phys.\ Rev.\ B {\bf 14}, 4274 (1976).\\
$[\mathrm{s}4]$ J C~Portal, R J~Nicholas, M A~Brummell, A Y~Cho, K Y~Cheng and T P~Pearsall,

\hspace{0.15cm}Solid\ State\ Commun. {\bf 43}, 907 (1982).\\
$[\mathrm{s}5]$ Th~Englert, J C~Maan, D C~Tsui and A C~Gossard, Solid\ State\ Commun. {\bf 45}, 989

\hspace{0.15cm}(1983).\\
$[\mathrm{s}6]$ F~Stern, Phys.\ Rev.\ B {\bf 5}, 4891 (1972).\\
$[\mathrm{s}7]$ T~Ando, A B~Fowler, and F~Stern, Rev.\ Mod.\ Phys. {\bf 54}, 437 (1982).\\
$[\mathrm{s}8]$ B~Gregory, J~Arthur, and G~Seidel, Phys.\ Rev.\ B {\bf 14}, 4274 (1976).\\
$[\mathrm{s}9]$ N Reyren, S Thiel, A D Caviglia, L~Fitting Kourkoutis, G Hammerl, C Richter, 

\hspace{0.15cm}C W Schneider, T Kopp, A-S R{\"u}etschi, D Jaccard, M Gabay, J-M Triscone, and 

\hspace{0.3cm}J Mannhart, Science {\bf 317}, 1196 (2007).\\
$[\mathrm{s}10]$ M~Basletic, J-L~Maurice, C~Carr{\'e}t{\'e}ro, G~Herranz, O~Copie, M~Bibes, {\'E}~Jacquet, 

\hspace{0.3cm}K~Bouzehouane, S~Fusil, and A~Barth{\'e}l{\'e}my, Nature Mater. {\bf 7}, 621 (2008).\\
$[\mathrm{s}11]$ A~Brinkman, M~Huijben, M~van Zalk, J~Huijben, U~Zeitler, J C~Maan, W G~van 

\hspace{0.3cm}der Wiel, G~Rijnders, D H A~Blank and H~Hilgenkamp, Nat.\ Mater. {\bf 6}, 493 (2007).\\
$[\mathrm{s}12]$ J A~Bert, B~Kalisky, C~Bell, M~Kim, Y~Hikita, H Y~Hwang and K~A~Moler, Nat.\ 

\hspace{0.3cm}Phys. {\bf 7}, 767 (2011).\\
$[\mathrm{s}13]$ M~Takizawa, S~Tsuda, T~Susaki, H Y~Hwang and A~Fujimori, Phys.\ Rev.\ B {\bf 84}, 

\hspace{0.3cm}245124 (2011).\\
$[\mathrm{s}14]$ Z S~Popovi{\'c}, S~Satpathy, and R M~Martin, Phys. Rev. Lett. {\bf 101}, 256801 (2008).\\
$[\mathrm{s}15]$ W-j~Son, E~Cho, B~Lee, J~Lee and S~Han, Phys.\ Rev.\ B {\bf 79}, 245411 (2009).\\

\end{document}